\documentclass[showpacs]{revtex4}
\usepackage{amsfonts}
\usepackage{amsmath}
\usepackage{latexsym}
\usepackage{epsfig}

\begin{document}

\title{Non-Markovian random walks and non-linear reactions: subdiffusion and
propagating fronts }
\author{Sergei Fedotov}
\affiliation{School of Mathematics, The University of Manchester, Manchester, M13 9PL, UK }

\begin{abstract}
The main aim of the paper is to incorporate the non-linear kinetic term into
non-Markovian transport equations described by a continuous time random walk (CTRW)
with non-exponential waiting time distributions. We consider three different
CTRW models with reactions. We derive new non-linear Master equations for
the mesoscopic density of reacting particles corresponding to CTRW with
arbitrary jump and waiting time distributions. We apply these equations to
the problem of front propagation in the reaction-transport systems with
Kolmogorov-Petrovskii-Piskunov (KPP) kinetics and anomalous diffusion.
We have found an explicit expression
for the speed of a propagating front in the case of subdiffusive transport.
\end{abstract}

\maketitle

\section{ Introduction}

This article addresses the problem of the mesoscopic description of
reaction-transport system of particles performing a continuous time random
walk (CTRW) \cite{MK}. One of the main challenges is an implementation of the
description of 
chemical reactions in non-Markovian transport processes governed by CTRW
with non-exponential waiting time distributions. There exist several
approaches and techniques to deal with this problem \cite{VR,FM,FO,
H,Seki,Sh,Nep,Meer,FM2,Shk}. In particular, there are many efforts to
incorporate the chemical reactions into subdiffusive transport. Different
models lead to various fractional reaction-diffusion equations corresponding
to kinetic regimes \cite{Sokolov, SSS, Froem,Henry,Hor,FI,E,Cam,Z} and
subdiffusion-limited reactions \cite{Yuste}.

Our main objective here is to discuss how to incorporate the\textit{\
non-linear} kinetic term into \textit{non-Markovian} transport equations which is
still an open problem. We consider a one-component reaction-transport system
consisting of independent particles $X$ that follow CTRW. Let $\rho \left(
x,t\right) $ represent the density of these particles at point $x$ and time $%
t$. The main purpose is to derive the non-linear Master equation for the
density $\rho (x,t)$ in the following form
\begin{equation}
\frac{\partial \rho }{\partial t}=L\rho ,  \label{MM}
\end{equation}%
where the non-linear evolution operator $L$ has to be determined. The
challenge is to derive the Master equation for an arbitrary CTRW model
coupled with non-linear reaction. We assume that the chemical reaction
follows the mass action law and the reaction term is of the form $r\left(
\rho \right) \rho .$ The density of other species participating in the
reactions are held constant. It has been shown recently that for
non-Markovian transport we cannot just add the term $r\left( \rho \right)
\rho $ to the right hand side of the evolution equation (\ref{MM}) \cite%
{Sokolov, SSS, Froem,Henry,Hor}. It is convenient to represent the
non-linear reaction rate $r\left( \rho \right) $ as the difference between
the birth rate $r^{+}\left( \rho \right) $ and the death rate $r^{-}\left(
\rho \right) $%
\begin{equation}
r\left( \rho \right) =r^{+}\left( \rho \right) -r^{-}\left( \rho \right) .
\end{equation}%
As an illustration, let us consider the classical Schl\"{o}gl first model
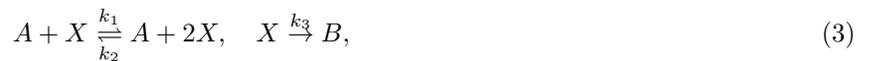
\cite{Halvin}
\begin{equation}
A+X\underset{k_{2}}{\overset{k_{1}}{\rightleftharpoons }}A+2X,\quad X\overset%
{k_{3}}{\rightarrow }B,  \label{sh1}
\end{equation}%
where $k_{1,\text{ }}k_{2}$ and $k_{3}$ are the reaction rate constants. If
we denote the densities of particles $X$ and $A$ by $\rho $ and $\rho _{A}$
respectively, then the birth and death rates are
\begin{equation}
r^{+}\left( \rho \right) =k_{1}\rho _{A},\quad r^{-}\left( \rho \right)
=k_{3}+k_{2}\rho .  \label{sh2}
\end{equation}%
In what follows the density $\rho _{A}$ of the catalyst $A$ is assumed to be
constant.

In what follows we consider three different non-Markovian models for
reaction and transport processes. We apply  these models to the problem of
propagating fronts in reaction-transport systems with non-standard diffusion
\cite{Fprl,Saar} (see also a recent review \cite{Fort}). The theory of
subdiffusive propagation of a front is presented in \cite{FM,C,V,Yadav,
Froemberg,Shk,CM}, the superdiffusive propagation is studied in \cite{Cast}.

\section{Model A}

Non-Markovian behavior of particles performing CTRW occurs when diffusive
particles get trapped for random times with non-exponential distribution.
Let $\phi \left( t\right) $ and $w\left( z\right) $ denote the waiting time
probability density function and the dispersal kernel respectively. For
simplicity we consider the uncoupled case when jumps and waiting times are
independent. If the distribution of waiting times is exponential: $\phi
\left( t\right) =\lambda \exp \left( -\lambda t\right) ,$ the transport
model is Markovian, and the difficulty of implementation of a non-linear
reaction term does not arise. In this case the Master equation (\ref{MM})
takes the form of the Kolmogorov-Feller equation with reaction
\begin{equation}
\frac{\partial \rho }{\partial t}=\lambda \int_{\mathbb{R}}\rho
(x-z,t)w\left( z\right) dz-\lambda \rho +r\left( \rho \right) \rho
\label{Master2}
\end{equation}%
with a clear separation of the transport and reactive terms on the RHS \cite%
{Fprl}. In the non-Markovian case these terms are not additive \cite%
{Sokolov, Henry,Hor,FI}. The key question is how the chemical reaction
influences the transport process. For Model A, we assume that transport
processes associated with CTRW and chemical reactions are independent. This
case has been considered in the series of papers \cite{Sokolov, SSS,Henry}.
The main assumption here is that when particles are trapped, the waiting
time is the same for all particles including newborn particles. One can
think of biological applications when cells or bacteria are trapped in some
confined region, say, at time $\tau <t$, they proliferate over the trapping
(waiting) time $t-\tau ,$ and then they are released at time $t$. In
particular, for the problem of virus infection and its propagation, the
random waiting time occurs due to virus reproduction inside infected cells
\cite{Fort1}.

\subsection{Conservation laws for particles}

The first step in the derivation of the non-Markovian Master equation (\ref%
{MM}) is to formulate the integral balance equations for the density $\rho
(x,t)$ and the auxiliary density $j\left( x,t\right) .$ The latter describes
the number of particles arriving at point $x$ \textit{exactly} at time $t$
(see, for example, \cite{Sokolov, FM2}). The balance equations for $\rho
\left( x,t\right) $ and $j\left( x,t\right) $ can be written in the
following form
\begin{equation}
\rho \left( x,t\right) =\rho _{0}\left( x\right) e^{\int_{0}^{t}r\left( \rho
\left( x,u\right) \right) du}\Psi \left( t\right) +\int_{0}^{t}j\left(
x,\tau \right) e^{\int_{\tau }^{t}r\left( \rho \left( x,u\right) \right)
du}\Psi \left( t-\tau \right) d\tau ,  \label{bal32}
\end{equation}%
\begin{eqnarray}
j\left( x,t\right) &=&\int_{\mathbb{R}}\rho _{0}\left( x-z\right)
e^{\int_{0}^{t}r\left( \rho \left( x,u\right) \right) du}w(z)\phi \left(
t\right) dz  \notag \\
&&+\int_{0}^{t}\int_{\mathbb{R}}j\left( x-z,\tau \right) e^{\int_{\tau
}^{t}r\left( \rho \left( x-z,u\right) \right) du}w(z)\phi \left( t-\tau
\right) dzd\tau .  \label{bal31}
\end{eqnarray}%
These equations represent the balance of particles due to non-linear
chemical reaction and transport process described by CTRW \cite{MK}. Eq. (%
\ref{bal32}) is the conservation law for particles' density $\rho \left(
x,t\right) $ at point $x$ at time $t$. The first term on the RHS of (\ref%
{bal32}) represents the particles that stay at their initial position up to
time $t$. Their density grows with the rate $r\left( \rho \left( x,t\right)
\right) $ during time interval $(0,t)$. The first term involves also the
survival function $\Psi \left( t\right) =\int_{t}^{\infty }\phi \left(
t\right) dt$ which is the probability that particles stay at their initial
location up to time $t$. The second term gives the number of particles that
arrive at point $x$ at previous time $\tau <t$ and grow with rate $r\left(
\rho \right) $ during time interval $\left( \tau ,t\right) $ so that no
jumps take place during this time interval. Note that the initial
distribution of particles $\rho _{0}\left( x\right) $ is set up in such a
way that a random walk for all particles starts at $t=0$ (no aging effects)
\cite{Bar}. Eq. (\ref{bal32}) is the conservation law for the particles that
arrive at point $x$ \textit{exactly} at time $t$. The first term on the RHS
of (\ref{bal31}) represents the particles that are at the point $x-z$ at
time $t=0$. Their density increases with the rate $r\left( \rho \right) $
during time interval $\left( 0,t\right) $ and they jump to the point $x$ at
time $t$. The second term describes the particles that arrive at the point $%
x-z$ at some time $\tau <t$ and react up to time $t$ at which the particles
jump to position $x.$ It should be noted that (\ref{bal32}) and (\ref{bal31}%
) are mesoscopic mean-field equations. We neglect the internal fluctuations
due to the finite number of particles. In general random fluctuations could
modify the macroscopic behavior of the reaction-transport systems (see, for
example, \cite{Gar,Lev}).

\subsection{Non-linear Master equation}

Let us now derive the evolution equation for the density $\rho \left(
x,t\right) $. Since the balance equations (\ref{bal32}) and (\ref{bal31})
are non-linear, we cannot apply directly the standard technique of
Fourier-Laplace transforms. Instead, we differentiate the density $\rho
\left( x,t\right) $ given by (\ref{bal32}) with respect to time
\begin{equation}
\frac{\partial \rho }{\partial t}=j\left( x,t\right) +r\left( \rho \right)
\rho -\rho _{0}\left( x\right) e^{\int_{0}^{t}r\left( \rho \left( x,u\right)
\right) du}\phi \left( t\right) -\int_{0}^{t}j\left( x,\tau \right)
e^{\int_{\tau }^{t}r\left( \rho \left( x,u\right) \right) du}\phi \left(
t-\tau \right) d\tau .  \label{bal33}
\end{equation}%
The last two terms can be interpreted as the density of particles that leave
the point $x$ exactly at time $t$
\begin{equation}
i\left( x,t\right) =\rho _{0}\left( x\right) e^{\int_{0}^{t}r\left( \rho
\left( x,u\right) \right) du}\phi \left( t\right) +\int_{0}^{t}j\left(
x,\tau \right) e^{\int_{\tau }^{t}r\left( \rho \left( x,u\right) \right)
du}\phi \left( t-\tau \right) d\tau .  \label{bal34}
\end{equation}%
It follows from (\ref{bal31}) and (\ref{bal34}) that%
\begin{equation}
j\left( x,t\right) =\int_{\mathbb{R}}i\left( x-z,t\right) w\left( z\right)
dz.
\end{equation}%
Then (\ref{bal33}) can be rewritten as
\begin{equation}
\frac{\partial \rho }{\partial t}=\int_{\mathbb{R}}i\left( x-z,t\right)
w\left( z\right) dz-i\left( x,t\right) +r\left( \rho \right) \rho .
\label{bal35}
\end{equation}%
This equation has a very simple meaning of a balance of particles at point $%
x.$ The first term on the RHS gives the number of particles coming to $x$
from different positions $x-z,$ where the jump size $z$ has the distribution
$w\left( z\right) $. The second term gives the rate at which the particles
leave the position $x$. The last term describes the growth of particles due
to chemical reactions. In the linear case, a similar equation has been used
\cite{Sokolov, SSS}. The advantage of having this equation is that we do not
need the Fourier transform to get the closed equation for the density $\rho
\left( x,t\right) .$ We can now find an expression for the density $i\left(
x,t\right) $ in terms of $\rho \left( x,t\right) $. We divide (\ref{bal32})
and (\ref{bal34}) by the factor $\exp \left( \int_{0}^{t}r\left( \rho \left(
x,u\right) \right) du\right) $ and take the Laplace transform of both
equations
\begin{equation*}
\mathcal{L}\left\{ \rho \left( x,t\right) e^{-\int_{0}^{t}r\left( \rho
\left( x,u\right) \right) du}\right\} =\left[ \rho _{0}\left( x\right) +%
\mathcal{L}\left\{ j\left( x,t\right) e^{-\int_{0}^{t}r\left( \rho \left(
x,u\right) \right) du}\right\} \right] \tilde{\Psi}\left( s\right) ,
\end{equation*}%
\begin{equation*}
\mathcal{L}\left\{ i\left( x,t\right) e^{-\int_{0}^{t}r\left( \rho \left(
x,u\right) \right) du}\right\} =\left[ \rho _{0}\left( x\right) +\mathcal{L}%
\left\{ j\left( x,t\right) e^{-\int_{0}^{t}r\left( \rho \left( x,u\right)
\right) du}\right\} \right] \tilde{\phi}\left( s\right) ,
\end{equation*}%
where $\tilde{\Psi}\left( s\right) =\mathcal{L}\left\{ \Psi \left( t\right)
\right\} $ and $\tilde{\phi}\left( s\right) =\mathcal{L}\left\{ \phi \left(
t\right) \right\} $. From these equations, we obtain
\begin{equation*}
\mathcal{L}\left\{ i\left( x,t\right) e^{-\int_{0}^{t}r\left( \rho \left(
x,u\right) \right) du}\right\} =\frac{\tilde{\phi}\left( s\right) }{\tilde{%
\Psi}\left( s\right) }\mathcal{L}\left\{ \rho \left( x,t\right)
e^{-\int_{0}^{t}r\left( \rho \left( x,u\right) \right) du}\right\} .
\end{equation*}%
Inverse Laplace transform gives
\begin{equation}
i\left( x,t\right) =\int_{0}^{t}K\left( t-\tau \right) \rho \left( x,\tau
\right) e^{\int_{\tau }^{t}r\left( \rho \left( x,u\right) \right) du}d\tau ,
\label{bal38}
\end{equation}%
where $K\left( t\right) $ is the standard memory kernel defined by its
Laplace transform
\begin{equation}
\tilde{K}\left( s\right) =\frac{\tilde{\phi}\left( s\right) }{\tilde{\Psi}%
\left( s\right) }=\frac{s\tilde{\phi}\left( s\right) }{1-\tilde{\phi}\left(
s\right) }.  \label{L}
\end{equation}%
Substitution of (\ref{bal38}) into (\ref{bal35}) gives us the non-linear
Master equation
\begin{eqnarray}
\frac{\partial \rho }{\partial t} &=&\int_{0}^{t}K\left( t-\tau \right) %
\left[ \int_{\mathbb{R}}\rho \left( x-z,\tau \right) e^{\int_{\tau
}^{t}r\left( \rho \left( x-z,u\right) \right) du}w\left( z\right) dz-\rho
\left( x,\tau \right) e^{\int_{\tau }^{t}r\left( \rho \left( x,u\right)
\right) du}\right] d\tau  \notag \\
&&+r\left( \rho \right) \rho .  \label{Master}
\end{eqnarray}%
This non-linear equation is the main result of this paper. For some
particular cases, it can be reduced to known equations in the literature.
For example, when the reaction rate $r\left( \rho \right) =r=const,$ Eq. (%
\ref{Master}) takes the form
\begin{equation}
\frac{\partial \rho }{\partial t}=\int_{0}^{t}K\left( t-\tau \right)
e^{r\left( t-\tau \right) }\left[ \int_{\mathbb{R}}\rho \left( x-z,\tau
\right) w\left( z\right) dz-\rho \left( x,\tau \right) \right] d\tau +r\rho ,
\label{linear}
\end{equation}%
and this model with constant rate $r$ has been formulated in \cite{Henry}.
Since the effective memory kernel $K\left( t-\tau \right) e^{r\left( t-\tau
\right) }$ depends on the reaction rate, it is tempting to conclude that
this equation describes the coupling of chemical reaction and transport. We
believe that this conclusion is misleading. In fact this equation describes
the complete decoupling of transport with memory effects and linear
reaction. To show this, let us make a substitution
\begin{equation}
\rho \left( x,t\right) =n\left( x,t\right) e^{rt}.
\end{equation}%
Then we obtain the Master equation for $n\left( x,t\right) $
\begin{equation}
\frac{\partial n}{\partial t}=\int_{0}^{t}K\left( t-\tau \right) \left[
\int_{\mathbb{R}}n\left( x-z,\tau \right) w\left( z\right) dz-n\left( x,\tau
\right) \right] d\tau
\end{equation}%
which is independent of the reaction and describes the transport of passive
particles. So we have a complete decoupling in which the density $\rho
\left( x,t\right) $ is the product of the density of passive particles $%
n\left( x,t\right) $ and the exponential factor $e^{rt}$ due to the chemical
reaction.

\subsection{Subdiffusive transport}

Now consider slow anomalous diffusion for which the waiting time pdf $\phi
(t)$\ has a power-law tail: $\phi (t)\sim (\tau _{0}/t)^{1+\gamma }$ with $%
0<\gamma <1$ as $t\rightarrow \infty .$ Clearly, the first moment $%
\int_{0}^{\infty }t\phi (t)dt$ is divergent for $0<\gamma <1.$ As a result,
the mean square displacement $<x^{2}>$ depends on time $t$ as $t^{\gamma }$
\cite{MK}. Here we use the following expression for the survival probability
\begin{equation}
\Psi \left( t\right) =E_{\gamma }\left[ -\left( \frac{t}{\tau _{0}}\right)
^{\gamma }\right] ,\quad 0<\gamma <1,  \label{ML}
\end{equation}%
where $E_{\gamma }\left[ x\right] =\sum_{0}^{\infty }x^{n}/\Gamma \left(
\gamma n+1\right) $ is the Mittag-Leffler function. The waiting time pdf $%
\phi (t)=-\frac{d}{dt}E_{\gamma }\left[ -\left( \frac{t}{\tau _{0}}\right)
^{\gamma }\right] $ has singular behavior $t^{\gamma -1}$as $t\rightarrow 0.$
The Laplace transforms of $\Psi \left( t\right) $ and $\phi (t)$ are
\begin{equation}
\tilde{\Psi}\left( s\right) =\frac{\tau _{0}\left( s\tau _{0}\right)
^{\gamma -1}}{1+\left( s\tau _{0}\right) ^{\gamma }},\quad \tilde{\phi}%
\left( s\right) =\frac{1}{1+\left( s\tau _{0}\right) ^{\gamma }}.
\end{equation}%
The advantage of using the Mittag-Leffler function is that we can find the
fractional reaction-transport equation without passing to the long-time
large-distance limit. We find from (\ref{L}) that the Laplace transform of
the memory kernel is%
\begin{equation}
\tilde{K}\left( s\right) =\frac{s^{1-\gamma }}{\tau _{0}^{\gamma }}.
\label{La}
\end{equation}%
Eq. (\ref{Master}) takes the form of a non-linear fractional equation
\begin{eqnarray}
\frac{\partial \rho }{\partial t} &=&\frac{e^{\int_{0}^{t}r\left( \rho
\left( x-z,u\right) \right) du}}{\tau _{0}^{\gamma }}D_{t}^{1-\gamma }\left(
\int_{\mathbb{R}}\rho \left( x-z,t\right) e^{-\int_{0}^{t}r\left( \rho
\left( x-z,u\right) \right) du}w\left( z\right) dz\right)  \notag \\
&&-\frac{e^{\int_{0}^{t}r\left( \rho \left( x,u\right) \right) du}}{\tau
_{0}^{\gamma }}D_{t}^{1-\gamma }\left( \rho \left( x,t\right)
e^{-\int_{0}^{t}r\left( \rho \left( x,u\right) \right) du}\right) +r\left(
\rho \right) \rho ,  \label{Master44}
\end{eqnarray}%
where $D_{t}^{1-\gamma }$ is the Riemann-Liouville fractional derivative
defined as
\begin{equation*}
D_{t}^{1-\gamma }\rho \left( x,t\right) =\frac{1}{\Gamma \left( 1-\gamma
\right) }\frac{\partial }{\partial t}\int_{0}^{t}\frac{\rho (x,\tau )d\tau }{%
\left( t-\tau \right) ^{\gamma }}.
\end{equation*}%
Now assume that the dispersal kernel $w\left( z\right) $ is an even and
rapidly decaying function for large $z$. We expand the expression in the
brackets (\ref{Master}) for small $z$ and truncate the Taylor series at the
second moment. We obtain
\begin{equation}
\frac{\partial \rho }{\partial t}=\frac{\sigma ^{2}}{2}\frac{\partial ^{2}}{%
\partial x^{2}}\int_{0}^{t}K\left( t-\tau \right) \rho \left( x,\tau \right)
e^{\int_{\tau }^{t}r\left( \rho \left( x,u\right) \right) du}d\tau +r\left(
\rho \right) \rho ,  \label{Master55}
\end{equation}%
where $\sigma ^{2}=\int_{\mathbb{R}}z^{2}w\left( z\right) dz$. Note that a
similar non-linear equation has been derived in \cite{Hor,E}.

By using the Laplace transform (\ref{La}) and the equation (\ref{Master55}),
we obtain the non-linear reaction-subdiffusion equation:
\begin{equation}
\frac{\partial \rho }{\partial t}=D\left( \gamma \right)
e^{\int_{0}^{t}r\left( \rho \left( x,u\right) \right) du}D_{t}^{1-\gamma }%
\frac{\partial ^{2}}{\partial x^{2}}\rho \left( x,t\right)
e^{-\int_{0}^{t}r\left( \rho \left( x,u\right) \right) du}+r\left( \rho
\right) \rho ,
\end{equation}%
where $D\left( \gamma \right) =$ $\sigma ^{2}/2\tau _{0}^{\gamma }$ is the
anomalous diffusivity.

\section{ Model B}

\textit{Model B }deals with the case when the transport process described by
CTRW depends on the chemical reaction. This model was considered by Vlad and
Ross \cite{VR}. They introduced the notion of the age of the particle as the
transition time between two successive jumps. The particles have zero age
when they just arrive at some point $x$ from which they will jump later. The
main assumption for Model B is that the newborn particles produced with the
rate $r^{+}\left( \rho \right) \rho $ have zero age. In other words, when a
new particle is born, it is given a new waiting time for a jump (zero age).
The density $j\left( x,t\right) $ for Model B can be interpreted as a
zero-age density of particles arriving at the point $x$ exactly at time $t.$
The particles arrive at the point $x$ because of the\ jumps in space and a
birth process with the rate $r^{+}\left( \rho \right) .$ Since the
non-linear reaction rate $r\left( \rho \right) $ is the difference between
the birth rate $r^{+}\left( \rho \right) $ and the death rate $r^{-}\left(
\rho \right) ,$ the balance equations for the densities $j\left( x,t\right) $
and $\rho \left( x,t\right) $ can be written as%
\begin{equation}
\rho \left( x,t\right) =\rho _{0}\left( x\right) e^{-\int_{0}^{t}r^{-}\left(
\rho \left( x,u\right) \right) du}\Psi \left( t\right) +\int_{0}^{t}j\left(
x,\tau \right) e^{-\int_{\tau }^{t}r^{-}\left( \rho \left( x,u\right)
\right) du}\Psi \left( t-\tau \right) d\tau  \label{b1}
\end{equation}%
and%
\begin{eqnarray}
j\left( x,t\right) &=&r^{+}\left( \rho \right) \rho +\int_{\mathbb{R}}\rho
_{0}\left( x-z\right) e^{-\int_{0}^{t}r^{-}\left( \rho \left( x,u\right)
\right) du}w(z)\phi \left( t\right) dz  \notag \\
&&+\int_{0}^{t}\int_{\mathbb{R}}j\left( x-z,\tau \right) e^{-\int_{\tau
}^{t}r^{-}\left( \rho \left( x,u\right) \right) du}w(z)\phi \left( t-\tau
\right) dzd\tau .  \label{b2}
\end{eqnarray}%
This system of equations has been derived in \cite{VR,Hor}.The authors
employed the Markov model with age-dependent density $\xi \left( x,t,\tau
\right) $ such that $\rho \left( x,t\right) =\int_{0}^{\infty }\xi \left(
x,t,\tau \right) d\tau $. Of course, (\ref{b1}) and (\ref{b2}) can be
formulated directly as the balance equations without introduction of $\xi
\left( x,t,\tau \right) $. \ Eq. (\ref{b1}) is the conservation law for the
particles at point $x$ at time $t$. Eq. (\ref{b2}) describes the situation
when the particles with zero age at point $x$ are either produced with a
rate $r^{+}\left( \rho \right) $ or arrive at point $x$ from other
positions.\ By using the method developed for Model A, one can derive the
non-linear Master equation for the density $\rho \left( x,t\right) $
\begin{eqnarray}
\frac{\partial \rho }{\partial t} &=&\int_{0}^{t}K\left( t-\tau \right) %
\left[ \int_{\mathbb{R}}\rho \left( x-z,\tau \right) e^{-\int_{\tau
}^{t}r^{-}\left( \rho \left( x-z,u\right) \right) du}w\left( z\right)
dz-\rho \left( x,\tau \right) e^{-\int_{\tau }^{t}r^{-}\left( \rho \left(
x,u\right) \right) du}\right] d\tau  \notag \\
&&+r^{+}\left( \rho \right) \rho -r^{-}\left( \rho \right) \rho .
\label{age}
\end{eqnarray}%
Note that Vlad and Ross stated that the balance equations (\ref{b1}) and (%
\ref{b2}) could not be reduced to a non-linear Master equation as (\ref{age}%
) due to the non-linear term $\exp \left( -\int_{\tau }^{t}r^{-}\left( \rho
\left( x-z,u\right) \right) du\right) $ \cite{VR}. However, Yadav and
Horsthemke managed to overcome this difficulty by using the large-spatial
scale and long-time limits \cite{Hor}. They used the standard asymptotics
for the Fourier transform of jump density $w\left( k\right) =1-\sigma
^{2}k^{2}+o(k^{2})$ and neglected the initial conditions in the long-time
limit. If we expand $\rho \left( x-z,\tau \right) $ in (\ref{age}) for small
$z$, then we obtain the equation derived in \cite{Hor}%
\begin{equation}
\frac{\partial \rho }{\partial t}=\frac{\sigma ^{2}}{2}\frac{\partial ^{2}}{%
\partial x^{2}}\int_{0}^{t}K\left( t-\tau \right) \rho \left( x,\tau \right)
e^{-\int_{\tau }^{t}r^{-}\left( \rho \left( x,u\right) \right) du}d\tau
+r^{+}\left( \rho \right) \rho -r^{-}\left( \rho \right) \rho .  \label{JHeq}
\end{equation}%
Thus, the Master equation (\ref{age}) can be considered as the
generalization of (\ref{JHeq}) derived in \cite{Hor} for the arbitrary jump
distribution $w\left( z\right) $.

The essence of the Model B and the main difference with Model A is that it
describes the situation when the newborn particles are given the new waiting
time for a jump. One can think of a situation in which the trapping
mechanism has been induced by chemical binding of newborn molecules. In this
case we have to take into account the aging effects. Of course, we should
make a clear distinction between the age of a jump event and the age of a
particle from $t=0$. The latter effect is not considered in this paper.

\section{Model C}

This model corresponds to the phenomenological generalization of the CTRW
model for the case when the chemical reaction is taken into account. We can
incorporate the local growth rate of diffusing particles by adding a term $%
\int_{0}^{t}r^{+}\left( \rho \left( x,t-\tau \right) \right) \rho \left(
x,t-\tau \right) \Psi \left( \tau \right) d\tau $ to the balance equation
for the density of particles. We write
\begin{eqnarray}
\rho \left( x,t\right) &=&\rho \left( x,0\right) \Psi \left( t\right)
+\int_{0}^{t}\int_{\mathbb{R}}\rho \left( x-z,t-\tau \right) w(z)\phi \left(
\tau \right) dzd\tau  \notag \\
&&+\int_{0}^{t}r^{+}\left( \rho \left( x,t-\tau \right) \right) \rho \left(
x,t-\tau \right) \Psi \left( \tau \right) d\tau .  \label{CTRWe}
\end{eqnarray}%
This equation has been used in \cite{FM,H} as a starting point. Henry,
Langlands and Wearne modified it in \cite{Henry} and pointed out that the
Model C is not justified at the mesoscopic level and what is more its
interpretation is not clear. The purpose of this section is to show that
Model C has a very natural physical interpretation.

Here we assume that the reaction is a pure birth process: $r^{+}\left( \rho
\right) ,$ and $j\left( x,t\right) $ is a zero-age density of particles
arriving at the point $x$ exactly at time $t$. The balance equations are
\begin{equation}
\rho \left( x,t\right) =\rho _{0}\left( x\right) \Psi \left( t\right)
+\int_{0}^{t}j\left( x,t-\tau \right) \Psi \left( \tau \right) d\tau
\label{modc1}
\end{equation}%
and%
\begin{eqnarray}
j\left( x,t\right) &=&r^{+}\left( \rho \right) \rho +\int_{\mathbb{R}}\rho
_{0}\left( x-z\right) w(z)\phi \left( t\right) dz  \notag \\
&&+\int_{0}^{t}\int_{\mathbb{R}}j\left( x-z,t-\tau \right) w(z)\phi \left(
\tau \right) dzd\tau .  \label{modc2}
\end{eqnarray}%
These equations do not involve non-linear terms inside the integrals.
Therefore, the standard technique of Fourier-Laplace transforms can be
employed to reduce the two balance equations (\ref{modc1}) and (\ref{modc2})
to a single equation for $\rho \left( x,t\right) $. It turns out that this
equation can be written as a phenomenological balance equation (\ref{CTRWe}%
). So Eq. (\ref{CTRWe}) corresponds to a mesoscopic situation when the
newborn particles are given a new waiting time for a jump. Note that it has
been found \cite{Henry} that we cannot use the balance equation (\ref{CTRWe}%
) with negative reaction term, since it leads to the negative density for
the sub-diffusive transport. Of course, Model C is just a particular case of
Model B when the death rate $r^{-}\left( \rho \right) =0.$

From (\ref{CTRWe}) one can obtain the Master equation for the density $\rho
\left( x,t\right) $ in the following form
\begin{equation}
\frac{\partial \rho }{\partial t}=\int_{0}^{t}K\left( t-\tau \right) \left[
\int_{\mathbb{R}}\rho \left( x-z,\tau \right) w\left( z\right) dz-\rho
\left( x,\tau \right) \right] d\tau +r^{+}\left( \rho \right) \rho  \notag
\end{equation}%
in which the transport term does not directly depend on the chemical
reaction as in Models A and B.

\section{Speed of traveling waves}

In this section we address the problem of wavefront propagation for Model A.
We assume that the reaction rate $r\left( \rho \right) $ is of the
Kolmogorov-Petrovskii-Piskunov (KPP) type \cite{FM2,Fr85}:
\begin{equation}
\max_{0\leq \rho \leq 1}r\left( \rho \right) =r\left( 0\right) >0,\qquad
r(1)=0.
\end{equation}%
Note that the standard logistic growth corresponds to $r\left( \rho \right)
=1-\rho $ \cite{Murray}. We start with the non-linear Master equation for
the density $\rho \left( x,t\right) $
\begin{equation}
\frac{\partial \rho }{\partial t}=\frac{\sigma ^{2}}{2}\frac{\partial ^{2}}{%
\partial x^{2}}\int_{0}^{t}K\left( t-\tau \right) \rho \left( x,\tau \right)
e^{\int_{\tau }^{t}r\left( \rho \left( x,u\right) \right) du}d\tau +r\left(
\rho \right) \rho  \label{main1}
\end{equation}%
with an initial condition in the form of a step function
\begin{equation}
\rho \left( x,0\right) =\theta \left( x\right) ,  \label{in}
\end{equation}%
where $\theta \left( x\right) =1$ for $x\leq 0$ and $\theta (x)=0$ for $x>0$%
. This condition describes the initial segregation of an unstable state $%
(\rho =0)$ for $x>0$ and a stable state $(\rho =1)$ for $x\leq 0.$

The purpose is to find the traveling wave solution $\rho \left( x,t\right)
=f\left( x-vt\right) $ of the initial value problem (\ref{main1}) and (\ref%
{in}). Here $v$ is the speed at which the wave profile $f$ invades the
unstable state with $\rho =0.$ When the memory kernel $K\left( t-\tau
\right) =\tau _{0}^{-1}\delta (t-\tau ),$ then Eq. (\ref{main1}) becomes the
KPP-equation (Fisher equation)%
\begin{equation}
\frac{\partial \rho }{\partial t}=D\frac{\partial ^{2}\rho }{\partial x^{2}}%
+r\left( \rho \right) \rho  \label{kpp}
\end{equation}%
with the diffusion coefficient $D=\sigma ^{2}/2\tau _{0}.$ This case
corresponds to the exponential waiting time pdf\ $\phi \left( t\right)
=\lambda \exp \left( -\lambda t\right) $ with $\lambda =\tau _{0}^{-1}.$ It
is well-known \cite{Murray} that the minimal propagation speed $v$ for (\ref%
{kpp}) is $2\sqrt{Dr\left( 0\right) }$. \

To find the propagation rate for the non-Markovian initial value problem (%
\ref{main1}) and (\ref{in}), we use the Hamilton-Jacobi approach \cite%
{Fprl,Fr85}. The starting point is to apply hyperbolic scaling $x\rightarrow
x/\varepsilon $, $t\rightarrow t/\varepsilon $ with $\varepsilon \rightarrow
0$ which corresponds to the long-time large-distance behavior of the
traveling wave. When the scaling parameter $\varepsilon \rightarrow 0,$ the
rescaled density $\rho ^{\varepsilon }(x,t)=\rho (x/\varepsilon
,t/\varepsilon )$ can take only two values $0$ and $1$ everywhere except in
the narrow front region where the transport and reaction terms are balanced.
In another words, the wave profile $f\left( \left( x-vt\right) /\varepsilon
\right) $ tends to a unit step function $\theta (x-vt).$ The aim is to find
the location of the front and rate at which it moves.

We introduce the action functional $G^{\varepsilon }$ as%
\begin{equation}
\rho ^{\varepsilon }(x,t)=\exp \left[ -\frac{G^{\varepsilon }(x,t)}{%
\varepsilon }\right] .  \label{an}
\end{equation}%
It follows from (\ref{an}) that if the function $G(x,t)=\lim_{\varepsilon
\rightarrow 0}G^{\varepsilon }(x,t)$ is positive, the rescaled density $\rho
^{\varepsilon }(x,t)\rightarrow 0$ as $\varepsilon \rightarrow 0.$ The
boundary of the set $G(x,t)>0$ can be regarded as the reaction front \cite%
{Fprl,Fr85}. Then, the front position $x(t)$ can be determined by the
equation $G(x(t),t)=0.$ Since we are interested in the leading edge of
the traveling wave ($\rho ^{\varepsilon }\approx 0$), we write the equation
for the rescaled density in the linear form%
\begin{equation}
\frac{\partial \rho ^{\varepsilon }}{\partial t}=\frac{\varepsilon \sigma
^{2}}{2}\frac{\partial ^{2}}{\partial x^{2}}\int_{0}^{t/\varepsilon }K\left(
\tau \right) e^{r(0)\tau }\rho ^{\varepsilon }\left( x,t-\varepsilon \tau
\right) d\tau +\frac{r\left( 0\right) \rho ^{\varepsilon }}{\varepsilon }.
\label{rescaled}
\end{equation}%
Substituting (\ref{an}) into (\ref{rescaled}) and taking the limit $%
\varepsilon \rightarrow 0,$ we obtain\
\begin{equation}
\frac{\partial G}{\partial t}+\frac{\sigma ^{2}}{2}\left( \frac{\partial G}{%
\partial x}\right) ^{2}\int_{0}^{\infty }K\left( \tau \right) e^{r(0)\tau
}e^{\tau \frac{\partial G}{\partial t}}d\tau +r(0)=0.  \label{s18}
\end{equation}%
This is the Hamilton-Jacobi equation for the action functional $G(x,t)$ (see
\cite{FM,Yadav,Fprl}). If we introduce the Hamiltonian $H=-\frac{\partial G}{%
\partial t}$ and the generalized momentum $p=\frac{\partial G}{\partial x},$
then equation (\ref{s18}) can be written as
\begin{equation}
H+\tilde{K}\left( H-r(0)\right) \frac{\sigma ^{2}p^{2}}{2}-r(0)=0,  \label{H}
\end{equation}%
where $\tilde{K}\left( s\right) $ is the Laplace transform of the memory
kernel $K\left( t\right) .$ This equation is different from the analogous
one in \cite{FM,Yadav}.

The propagation rate $v$ can be found from \cite{Fprl}
\begin{equation}
v=\frac{\partial H}{\partial p}=\frac{H}{p}.  \label{Ha1}
\end{equation}%
Solving quadratic equation (\ref{H}) for $p,$ we obtain from (\ref{Ha1})
that
\begin{equation}
v=H^{\ast }\sigma \sqrt{\frac{\tilde{K}\left( H^{\ast }-r(0)\right) }{%
2H^{\ast }-2r(0)}},  \label{velo}
\end{equation}%
where $H^{\ast }$ is the solution of
\begin{equation}
\frac{\partial }{\partial H}\left( \frac{H-r(0)}{\tilde{K}\left( H-r\right)
(0)}\right) =\frac{2\left( H-r(0)\right) }{H\tilde{K}\left( H-r(0)\right) }.
\label{s21}
\end{equation}

\subsection{Front propagation rate for subdiffusive transport}

Let us consider the reaction-subdiffusion case for which the Laplace
transform of the waiting time pdf $\phi \left( t\right) $ is%
\begin{equation*}
\tilde{\phi}\left( s\right) =\frac{1}{1+\left( s\tau _{0}\right) ^{\gamma }}%
,\qquad 0<\gamma <1
\end{equation*}%
and the Laplace transform of the memory kernel is given by (\ref{La}). Then
Eq. (\ref{s21}) has the solution $H^{\ast }=2r(0)/\left( 2-\gamma \right) .$
Substitution of this solution into (\ref{velo}) gives the propagation rate $v
$ corresponding to the Master equation (\ref{main1}) with the initial
condition (\ref{in}). We introduce the following notations for the
propagation speeds: $v_{A}$ for Model A;\ $v_{B}$ and $v_{C}$ for Model B
and Model C respectively. The minimal propagation speed $v_{A}$ for Model A
with (\ref{main1}) and (\ref{in}) is
\begin{equation}
v_{A}=\sqrt{\frac{2r(0)^{2-\gamma }\sigma ^{2}}{\tau _{0}^{\gamma }\left(
2-\gamma \right) ^{2-\gamma }\gamma ^{\gamma }}}.  \label{fin}
\end{equation}%
For simplicity, we consider the case when the death rate $r^{-}\left( \rho
\right) $ obeys $r^{-}\left( 0\right) =0.$ For the Schl\"{o}gl first model (%
\ref{sh1}), it means that $k_{3}=0$ (see (\ref{sh2}). Then for Model B and
Model C we have \cite{FM,Yadav}
\begin{equation}
v_{B}=v_{C}=\sqrt{\frac{r(0)^{2-\gamma }\sigma ^{2}\left( 3-\gamma \right)
^{3-\gamma }}{2\tau _{0}^{\gamma }\left( 2-\gamma \right) ^{2-\gamma }}}.
\end{equation}%
The case $k_{3}\neq 0$ was considered in \cite{Yadav}.

For $\gamma =1,$ we have the classical result $v=2\sqrt{Dr(0)}$ with $%
D=\sigma ^{2}/2\tau _{0}$ that corresponds to the KPP-equation (Fisher
equation) (\ref{kpp}). For $0<\gamma <1,$ the propagation speed $v_{B}$ is
greater than $v_{A}$. This is because the newborn particles in Model B and
Model C are given new waiting times for the jump event. As a result the
overall transport process and the propagation rate for Model A are slower
than those of Model B and Model C. It is also interesting to compare our
results with those obtained in \cite{Froemberg}. The authors considered the
irreversible autocatalytic reaction $A+X\rightarrow 2X$ with subdiffusion.
They found that the minimal propagation is zero. This finding seems to
contradict with our result of finite speed propagation (\ref{fin}). In fact,
Model A and the subdiffusion-reaction model studied in \cite{Froemberg} are
different. In our model we keep the concentration of one of the component
constant, while Froemberg \textit{et al }considered\textit{\ }two-component
system of equations for which both reactants $A$ and $X$ vary in space and
time. For example, in our paper the density $\rho _{A}$ of the catalyst $A$
in (\ref{sh1}) is assumed to be constant.

\section{ Conclusions}

In this paper we have given a mesoscopic description of a reaction-transport
system of particles performing a continuous time random walk (CTRW) with
non-exponential waiting time distributions. Our main objective has been to
implement a \textit{non-linear} kinetic term into \textit{non-Markovian}
transport equations. We have considered three different CTRW models with reactions
which have been discussed in the literature. We have derived new non-linear
Master equations (\ref{Master}) and (\ref{age}) for the mesoscopic density
of reacting particles corresponding to CTRW with arbitrary jump and waiting
time distributions. We have applied the theory to the problem of front
propagation in reaction-transport systems with KPP-kinetics and
non-Markovian diffusion. We have found an explicit expression for the speed
of the propagating front in the case of subdiffusion transport.

\textbf{Acknowledgement}. Author acknowledges the warm hospitality of the
Grup de F\'{\i}sica Estad\'{\i}stica, Universitat Aut\'{o}noma de Barcelona
and discussions with D. Campos and V. M\'{e}ndez. It is a pleasure to 
acknowledge discussions with Werner Horsthemke. This work has been
supported by the grants NEST-028192-FEPRE and 2007PIV-0001.

\end{document}